\documentclass{article}
\usepackage{amsmath}
\usepackage{amssymb}
\usepackage{graphicx}
\begin{document}
\begin{center}
\textbf{On the homogeneity of the
quantum transition probability}
\end{center}
\begin{center}
Gerd Niestegge
\footnotesize
\vspace{0,2 cm}

Ahaus, Germany

gerd.niestegge@web.de, https://orcid.org/0000-0002-3405-9356
\end{center}
\normalsize
\begin{abstract}
In the years 1952 and 1965, H.-C. Wang
and U. Hirzebruch showed that 
the two-point homogeneous compact spaces with convex metrics
are isometric to the spheres,
the real, complex, octonion projective spaces
and the Moufang plane 
and as well to 
the sets of the minimal idempotents or pure states
in the simple Euclidean Jordan algebras.
Here we reveal the physical meaning 
of these mathematical achievements for the
quantum mechanical transition probability.
We show that this transition probability
features a maximum degree of homogeneity
in all simple Euclidean Jordan algebras,
which includes common finite-dimensional Hilbert space quantum theory. 
The atomic parts of these algebras
or, equivalently, the extreme boundaries of 
their state spaces can be characterized by purely topological means. 
This is an important difference 
to many other recent approaches 
that aim to distinguish the entire state
spaces among the convex compact sets.
An interesting case with non-homogeneous transition probability arises,
when the $E_6$-symmetric bioctonionic projective plane
is used as quantum logic.\\

\end{abstract}

\section{Introduction}

A important mathematical accomplishment \cite{hirzebruch1965, wang1952two-pointhom} is
that the following objects are \emph{two-point homogeneous} compact spaces with \emph{convex metrics}
and that any two-point homogeneous compact space with a convex metric
is isometric to one of them:
the spheres, the real, complex, octonion projective spaces and the Moufang plane.
On the other hand, these objects coincide with the atomic parts 
(or, equivalently, with the extreme boundaries of the state spaces - the pure states) 
of the finite-dimensional simple \emph{Euclidean Jordan algebras}
(which are also called \emph{formally real} instead of Euclidean).
These are the spin  factors, the Hermitian $n \times n$-matrix algebras ($n = 3,4,5,... $) over 
the real numbers ($\mathbb{R}$), complex numbers ($\mathbb{C}$), 
quaternions ($\mathbb{H}$), and the Hermitian 
$3 \times 3$-matrix algebra over the octonions ($\mathbb{O}$), 
equipped with the Jordan product.
The complex matrix algebras form the usual model 
of finite-dimensional quantum theory.

Here we reveal the meaning of these well-established 
mathematical achievements for the
quantum mechanical transition probability. 
By Hirzebruch's results, we get that this transition probability
features a maximum degree of homogeneity
in all simple Euclidean Jordan algebras.
A theorem by Wang then characterizes
the atomic parts of these algebras
or, equivalently, the extreme boundaries of 
their state spaces (the pure state spaces) 
by purely topological means. 
This is an important difference 
to many other more recent approaches 
that aim to distinguish the entire state
spaces among the convex compact sets~\cite{AS02, 
BarnumHilgert2020, mielnik1968geometry, 
muller2012ududec, nie2023conv_self-dual}.

In the non-Abelian non-simple (reducible) cases,
the transition probability becomes non-homogeneous.
A further more interesting non-homogeneous case arises, when the $E_6$-symmetric 
bioctonionic projective plane \cite{atsuyama1985E6, baez2002octonions, 
catto2003, Corradetti2022conj-matters, gursey1978octonionic}
is used as quantum logic.

Section 2 contains a very brief synopsis of the 
theory of the Euclidean Jordan algebras and
focuses on those issues of the theory
that will be needed in this paper. 
In section 3 the transition probability 
and the homogeneity feature are treated.
The relation between this feature 
and the two-point homogeneity of metric spaces
is studied in section 4, where our main results 
are presented. Non-homogeneous
transition probability is considered in section 5.

\section{Euclidean Jordan algebras}

The Euclidean Jordan algebras are identical with the 
formally real Jordan algebras and the 
finite-dimensional JB algebras, although 
these structures are defined in three slightly different 
ways \cite{AS02, faraut1994analysis, hanche1984jordan}.
The common part of all definitions is the structure 
of a unital algebra over the real numbers 
with a commutative, but not associative product $\circ$ 
satisfying the Jordan identity 
$a \circ (a^{2} \circ b) = a^{2} \circ (a \circ b)$. 
Here the unit element is denoted by $\mathbb{I}$.
A Euclidean Jordan algebra is called \emph{simple} or \emph{irreducible},
when it cannot be represented as a direct sum of subalgebras.
The simple Euclidean Jordan algebras are
the spin factors, the Hermitian $n \times n$-matrix algebras 
(equipped with the Jordan product)
over $\mathbb{R}$, $\mathbb{C}$, $\mathbb{H}$ and,
only in the case $n=3$, over $\mathbb{O}$ \cite{AS02, hanche1984jordan}.
The matrix algebras are denoted by $H_n(\mathbb{R})$, $H_n(\mathbb{C})$, 
$H_n(\mathbb{H})$ and $H_3(\mathbb{O})$.

Any Euclidean Jordan algebra $A$ possesses an order relation $\leq$,
where we have $0 \leq a$ iff $a \in \left\{x^{2}| x \in A \right\}$.
A linear functional $\mu : A \rightarrow \mathbb{R}$ is \emph{positive},
if $0 \leq \mu(x^{2})$ for all $x \in A$, and $\mu$ is a state,
if $\mu$ is positive and $\mu(\mathbb{I})=1$ holds.
The state space shall be denoted by $S_A$. 

The idempotent elements in $A$ form an \emph{orthomodular lattice}, also
called the\textit{ quantum logic} $L_A$. A pair $p,q \in L_A$ is
orthogonal if $p + q \leq \mathbb{I}$ and $p' := \mathbb{I} - p $
becomes the orthocomplement of $p$. 
The minimal non-zero elements in $L_A$ are called \emph{atoms} 
or \emph{primitive elements}; $E_A$ denotes the set containing the atoms.
For each atom $p \in E_A$ there is a unique state $\mu \in S_A $ 
with $\mu(p)=1$. This state is an extreme point of $S_A$; this means 
that $\mu$ is a \emph{pure state}.

An automorphism of $A$ is a an invertible positive transformation 
$T: A \rightarrow A$ with a positive inverse and $T(\mathbb{I}) = \mathbb{I}$. The 
automorphisms form a group denoted by $Aut(A)$. This becomes a compact group, 
since the dimension of $A$ is finite.

$A$ possesses a \emph{trace}; this is a positive linear functional 
that is invariant under all automorphisms and satisfies $trace(p) = 1$
for the atoms $p \in E_A$. Moreover, $A \ni a \rightarrow trace(p \circ a)$
is the unique state $\mu$ with $\mu(p)=1$.
\newpage

With any $p \in L_A$ the subspace $\left\{p,A,p\right\}$
becomes a subalgebra with unit element $p$.
If $p$ is an atom, then $\left\{p,A,p\right\} = \mathbb{R}p$.
Here we use the so-called 
triple product which is defined as follows:
$\left\{ x,y,x \right\} := 2 x \circ ( x \circ y ) - x^{2} \circ y $ 
for $x,y \in A$.

If $A$ is simple and the atoms $p,q \in E_A$ are orthogonal,
they become \emph{strongly connected} \cite{AS02, hanche1984jordan}; this means that
there is an element $$v \in \left\{p,A,q\right\} \subseteq \left\{p+q,A,p+q\right\} $$ 
with $v^{2} = p + q$ ($v$ is a so-called \emph{symmetry}).
Then $a := (v + p + q)/2$ is idempotent with $a \leq p+q$, $0 \neq a \neq p+q$ 
and $p \neq a \neq q$. Thus $a$ is an atom.\\

\textbf{Lemma 1:} \textit{For any two atoms $p$ and $q$
in a simple Euclidean Jordan algebra $A$ 
there is a third atom $e$ with }
$$\arccos \sqrt{trace(p \circ q)} 
= \arccos \sqrt{trace(q \circ e)} + \arccos \sqrt{trace(e \circ p)}.$$

Proof. With $u := p \vee q$ we consider the subalgebra $\left\{u,A,u\right\}$
where $u$ becomes the unit element. With
$p' := u - p$ we have
$2 poq = p + \left\{qpq\right\} - \left\{q'pq'\right\} 
= p + sq - tq' $ with some $s,t \in \mathbb{R}$
and the product $p\circ q$ lies in the liner hull of 
the three elements $u,p,q$. Since these elements 
are idempotent and $u \circ p = p$, $u \circ q = q$,
they generate a Jordan subalgebra $A_o$
of $\left\{u,A,u\right\}$, 
the dimension which cannot exceed $3$. 
In the case $dim(A_o)=1$ we have $p=q$ and we can choose $e=p=q$.

We first consider the case $dim(A_o)=3$.
If $A_o = \mathbb{R} \oplus \mathbb{R} \oplus \mathbb{R}$,
the unit element $u$ cannot become the supremum of two atoms.
Therefore $A_o$ must be
isomorphic to the algebra consisting of the 
Hermitian real 2$\times$2-matrices
equipped with the Jordan product.
The atoms in $H_2(\mathbb{R})$ are the matrices 
$$ e_t := 
\begin{pmatrix}
cos^{2}(t) & - sin(t) cos(t)\\
- sin(t) cos(t)& sin^{2}(t)
\end{pmatrix}
$$
with $t \in \mathbb{R}$. Then
$$ trace(e_s \circ e_t) 
= \left[cos(s) cos(t) + sin(s) sin(t)\right]^{2} 
= cos^{2}(s - t) $$
with $s,t \in \mathbb{R}$. For $p = e_s$ and $q = e_r$ 
the atom $e := e_r$ with  $r := (s + t )/2$ then satisfies the above identity.

We now come to the case $dim(A_o)=2$, where 
$p$ and $q$ are orthogonal. 
Since $A$ is simple, $\left\{ u,A,u \right\}$ 
contains a further atom $a$
with $p \neq a \neq q$.
If the atoms $p$ and $a$ were orthogonal, we would get
$a= u - p = q$, which is ruled out. Therefore
$u, p, a$ generate a three-dimensional 
subalgebra containing $q = u - p$, 
and we can proceed as in the case $dim(A_o)=3$.
\hfill $\square$ \\

The set $E_A$ in any simple Euclidean Jordan algebra $A$ 
forms a Riemannian manifold and Lemma 1 also follows from the fact 
that $$d(p,q) = \arccos \sqrt{trace(p \circ q)}, \  p,q \in E_A,$$ 
is the geodesic distance on this manifold.

\section{Transition probability}

With a quantum logic $L$ and a state space $S$ the 
\emph{transition probability} from $p \in L$ to $q \in L$ exists
and becomes $s$ with $0 \leq s \leq 1$, if
$$ \left\{ \mu \in S | \mu(p)=1 \right\} \subseteq \left\{ \mu \in S | \mu(q)=s \right\}$$
holds \cite{nie2020alg_origin, nie2021generic}; 
it is denoted by $\mathbb{P}(q|p) := s$.

With the quantum logic $L_A$ and the state space $S_A$ 
of any Euclidean Jordan algebra $A$, the transition probability exists
for the elements $p$ and $q$ in $L_A$ iff
$\left\{ p,q,p \right\} = sp$ holds with some $s \in \mathbb{R}$;
in this case $\mathbb{P}(q|p) = s$ \cite{nie2021generic}.
When $p$ is an atom, $\mathbb{P}(q|p)$ exists for all $q \in L_A$ and
$\mathbb{P}(q|p) = trace(\left\{p,q,p\right\}) = trace(p \circ q)$. 

An automorphism $T \in Aut(A)$ maps $L_A$ to $L_A$ and $E_A$ to $E_A$. Moreover,
with any state $\mu \in S_A$,  the map
$A \ni a \rightarrow \mu (Ta)$ defines another state in $S_A$. Therefore the 
transition probability becomes invariant under the transformations $T \in Aut(A)$:
$$\mathbb{P}(q|p) = \mathbb{P}(Tq|Tp)$$
for all $p,q \in E_A$ and the maximum degree of homogeneity we can expect for the
transition probability is the following condition:
\begin{itemize}
\item[]
\textit{Whenever $\mathbb{P}(q_1|p_1) = \mathbb{P}(q_2|p_2)$ holds 
for two pairs of atoms $p_1,q_1$ and $p_2,q_2$, there is a 
transformation $T \in Aut(A)$ with $Tp_1 = p_2$ and $Tq_1 = q_2$.}
\end{itemize}
We call the transition probability \emph{homogeneous}, 
if this condition is satisfied. 

Another eased homogeneity condition is the so-called 
\emph{bit symmetry} \cite{muller2012ududec},
which postulates the existence of $T \in Aut(A)$ with $Tp_1 = p_2$ and $Tq_1 = q_2$
for atoms  $p_1, q_1, p_2, q_2$
only in the case $\mathbb{P}(q_1|p_1) = 0 = \mathbb{P}(q_2|p_2)$.

The Abelian Euclidean Jordan algebras have the form
$\mathbb{R} e_1 \oplus ... \oplus \mathbb{R} e_n$
with the atoms $e_1, ..., e_n$. 
They represent the finite classical probability spaces.
The automorphisms 
are the permutations of the atoms and 
the above homogeneity conditions
become trivial.

Though the homogeneity of the transition probability is a maximum requirement, 
it is also fulfilled in common Hilbert space quantum theory,
where the algebra $A$ consists of the bounded self-adjoint linear operators 
and atoms $p, q$ become projections on one-dimensional subspaces. 
In Dirac notation we get
$p = | \varphi >< \varphi |$ and $p = | \psi >< \psi |$
with normed vectors $ \varphi, \psi$ in the Hilbert space.
The transition probability is 
$\mathbb{P}(q|p) = trace(pq)
= \left|\left\langle \varphi | \psi \right\rangle\right|^{2}$.
This is the square of the cosinus of the angle 
between the vectors $ \varphi $ and $ \psi $.
With two further atoms 
$p_o = | \varphi_o >< \varphi_o |$ and $p_o = | \psi_o >< \psi_o |$
with the normed vectors $ \varphi_o, \psi_o$ 
and $\mathbb{P}(q|p) = \mathbb{P}(q_o|p_o)$
we have that the angle $\alpha$, $0 \leq \alpha < \pi$, 
between $ \varphi $ and $ \psi $ is identical 
with the angle between $ \varphi_o $ and $ \psi_o $.
Now we select a unitary transformation $u_1$ 
carrying $ \varphi $ to $ \varphi_o $.
The angle between $\varphi_o$ and $u_1 \psi $ is $\alpha$
and identical with the one between $\varphi_o$ and $\psi_o$.
Therefore there is a rotation $u_2$
around $ \varphi_o $ that carries $u_1 \psi $ to $\psi_o$.
Then $u_2 u_1 \varphi = \varphi_o$ and 
$u_2 u_1 \psi = \psi_o$. The desired transformation $T$
becomes the map $Ta := u_2 u_1 a u^{-1}_{1} u^{-1}_{2}$ 
for the self-adjoint operators~$a$.

In the next section we shall identify
further cases with homogeneous transition probability.

\section{Two-point homogeneous metric spaces}

A \emph{metric} $d$ on a space $X$ is called \emph{convex},
if there is a third point $x_o \in X$ with 
$$d(x_1, x_2) = d(x_1, x_o) + d(x_o, x_2)$$
for any pair $x_1, x_2 \in X$ \cite{wang1952two-pointhom}.
This means that the triangle inequality hold with equality
for the \emph{between point} $x_o$. A metric $d$ 
on a compact space $X$ can always be normalized 
in such a way that $sup \left\{ d(x,y)| x,y \in X \right\} = 1$.

A metric space $X$ with the metric $d$ is called 
\emph{two-point homogeneous} \cite{birkhoff1944metric, hirzebruch1965, wang1952two-pointhom}, 
if there is an isometry of the space carrying $a_1$ to $b_1$ and $a_2$ to $b_2$
for any two pairs of points $a_1, b_1$ and $a_2, b_2$ of the space with 
equal distances $d(a_1,b_1) = d(a_2,b_2)$. 

We shall now see 
how the two-point homogeneity is connected with the 
homogeneity of the transmission probability and
identify many cases with 
homogeneous transition probability, including 
the finite-dimensional version of common quantum theory.\\

\textbf{Theorem 1 (Hirzebruch 1965):}
\begin{itshape}
The atomic part $E_A$ of any simple Euclidean Jordan algebra $A$
becomes a two-point homogeneous compact metric set
with the normalized convex metric 
$$d(p,q) := \frac{2}{\pi} arccos \left(\sqrt{trace(p \circ q)}\right) 
= \frac{2}{\pi} arccos \left(\sqrt{\mathbb{P}(q|p)}\right)$$
for $p,q \in E_A$ and the
transition probability is homogeneous.\\
\end{itshape}

Proof. The Korollar following Satz 2.3 in \cite{hirzebruch1965} 
says that $(E_A, d)$ is a two-point homogeneous compact metric set. It is obvious
that the metric $d$ is normalized; its convexity follows from our Lemma 1.

Now suppose that $\mathbb{P}(q_1|p_1) = \mathbb{P}(q_2|p_2)$ holds 
for two pairs of atoms $p_1,q_1$ and $p_2,q_2$ in $E_A$. 
Since we have $\mathbb{P}(q|p) = trace(p \circ q)$ for $p,q \in E_A$, 
we get from \cite{hirzebruch1965} Satz 2.3 a transformation $T \in Aut(A)$
with $Tp_1 = p_2$ and $Tq_1 = q_2$ ($T$ is a combination of reflections).
\hfill $\square$\\

Two-point homogeneous compact sets with a convex metric 
are a rather abstract topological construct and it may surprise
that they all result in a certain quantum models with homogeneous transition 
probability, as we shall see now.\\

\textbf{Theorem 2 (Wang 1952):}
\begin{itshape} 
\textit
Let $(X, d)$ be a compact metric space with
the normalized metric $d$. 
If $(X, d)$ is two-point homogeneous 
and if the metric $d$ is convex, there is an isometry 
$\tau$ from $X$ to the atomic part $E_A$ of 
some simple Euclidean Jordan algebra A
with $$cos^{2}\left(\frac{\pi}{2}d(x,y)\right) 
= trace((\tau x) \circ (\tau y)) 
= \mathbb{P}(\tau x|\tau y)$$
for $x,y \in X$. \\
\end{itshape}

Proof. This is Wang's Theorem VI \cite{wang1952two-pointhom}. Normalizing
the metric, which he introduces for $E_A$ in \cite{wang1952two-pointhom} section 7, 
results in 
$$ \frac{2}{\pi} arccos \left(\sqrt{trace(p \circ q)}\right)$$
for $p,q \in E_A$, which gives us the above formula.
\hfill $\square$ \\

The maximum number of pairwise orthogonal atoms is $2$ with the spin factors
and is $n$ with the n$\times$n-matrix algebras. With the corresponding metric spaces $X$
this becomes the maximum number of elements with maximal pairwise distance. 
This maximal distance becomes $1$ with a normalized metric.
An unsolved problem, however, is to identify properties of the metric spaces 
that reveal with which on of the four number fields 
($\mathbb{R}$, $\mathbb{C}$, $\mathbb{H}$, $\mathbb{O}$) 
the corresponding matrix algebra is constructed.

\section{Non-homogeneous transition probability}

Examples with non-homogeneous transition probabilities
are immediately at hand. Consider any non-trivial simple 
Euclidean Jordan algebra $A_1 \neq \mathbb{R}$ 
and the direct sum $A := A_1 \oplus A_2$
with any further Euclidean Jordan algebra $A_2$.
Now select any two orthogonal atoms $p$ and $q_1$ in $A_1$
and a further atom $q_2 \in A_2$.
Then $\mathbb{P}(q_1|p) = 0 = \mathbb{P}(q_2|p)$. However,
a transformation $T \in Aut(A)$ with
$Tp = p$ and $Tq_1 = q_2$ cannot exist, since
$Tp \in A_1$ implicates $T(A_1) = A_1$.

We shall now address the more difficult task
to construct an irreducible example
with non-homogeneous transition probability.
This will become the bioctonionic projective plane
$(\mathbb{C} \otimes \mathbb{O})\mathbb{P}^{2}$,
which is the compact symmetric space
with the Cartan label $EIII$ \cite{atsuyama1985E6}.
Its isometry group is 
the exceptional Lie group~$E_6$~\cite{baez2002octonions}.

Note that this common terminology is
misleading, since $(\mathbb{C} \otimes \mathbb{O})\mathbb{P}^{2}$ is not 
a projective plane in the usual sense, but only in a wider sense
\cite{atsuyama1985E6, baez2002octonions, Corradetti2022conj-matters, gursey1978octonionic}. 
A representation can be constructed using the 
simple Jordan algebra $H_3(\mathbb{C} \otimes \mathbb{O})$.
These are the 3$\times$3-matrices with entries from $ \mathbb{C} \otimes \mathbb{O} $
that are Hermitian with respect to the octonion conjugation. The entries
in the diagonal become complex numbers.
There is a $\mathbb{R}$-valued positive-definite inner product 
$\left\langle x | y  \right\rangle := trace \left( x \circ y^{\ast} + y \circ x^{\ast} \right)/2$ 
on $H_3(\mathbb{C} \otimes \mathbb{O})$ \cite{gursey1978octonionic},
but $H_3(\mathbb{C} \otimes \mathbb{O})$ is not a Euclidean (formally real) Jordan algebra;
here $x^{\ast}$ denotes the complex conjugate transposed matrix.
The Freudenthal product $\times$ is defined 
in the following way for $x \in H_3(\mathbb{C} \otimes \mathbb{O})$:
$$ x \times x := x^{2} - x \  trace(x) - \frac{1}{2} \left[trace(x^{2}) - \left(trace(x)\right)^{2} \right] I,$$
where $I$ denotes the matrix with the entries $1$ in the diagonal 
and $0$ else \cite{catto2003, gursey1978octonionic}. Then 
$$ (\mathbb{C} \otimes \mathbb{O})\mathbb{P}^{2} 
= \left\{ p \in H_3(\mathbb{C} \otimes \mathbb{O}) | \left\langle p|p\right\rangle = 1 \text{ and } p \times p = 0 \right\} /\sim$$
with the following equivalence relation: 
$p_1 \sim p_2$ $:\Leftrightarrow $ $ p_1 = p_2$ or  $ p_1 = - p_2$.
A point in $(\mathbb{C} \otimes \mathbb{O})\mathbb{P}^{2}$ is the 
equivalence class $\bar{p} := \left\{p,-p\right\}$.
Unfortunately this equivalence relation and the construction of the 
equivalence classes is often suppressed in the literature.
Another mathematically rigorous way is to define the points of 
$(\mathbb{C} \otimes \mathbb{O})\mathbb{P}^{2}$ as the 
one-dimensional linear subspaces $\mathbb{R} p$ 
in $H_3(\mathbb{C} \otimes \mathbb{O})$ with 
$p \in H_3(\mathbb{C} \otimes \mathbb{O})$ and $ p \times p = 0$ 
and to select normalized elements in $\mathbb{R} p$ as representatives
\cite{Corradetti2022conj-matters}. 
For each point $\bar{p} \in (\mathbb{C} \otimes \mathbb{O})\mathbb{P}^{2}$ the set
$$\bar{p}^{\bot} := \left\{ \bar{q} \in (\mathbb{C} \otimes \mathbb{O})\mathbb{P}^{2} | \left\langle p|q\right\rangle = 0 \right\}$$
is a line in this geometry and for each line $l$ there is a unique point $\bar{p}$ such that
$l = \bar{p}^{\bot}$~\cite{Corradetti2022conj-matters}. 

Our quantum logic $L$ shall now consist of the points, the lines, the empty set and
the complete set $(\mathbb{C} \otimes \mathbb{O})\mathbb{P}^{2}$. The points become the atoms,
the empty set represents the $0$-element and 
the complete set $(\mathbb{C} \otimes \mathbb{O})\mathbb{P}^{2}$ 
represents the $\mathbb{I}$-element in the quantum logic $L$.
As we do not have a true projective plane~\cite{atsuyama1985E6}, this quantum logic 
is not a lattice, but only a partially ordered set (by set inclusion)
with an orthocomplementation that
maps an atom $\bar{p}$ to $\bar{p}^{\bot}$, a line $\bar{p}^{\bot}$ to the atom $\bar{p}$
and of course $0$ to $\mathbb{I}$ and $\mathbb{I}$ to $0$.
Note that the atoms, points and lines are not elements 
in the Jordan algebra $H_3(\mathbb{C} \otimes \mathbb{O})$,
the quantum logic is not a subset of it
and, therefore, its elements cannot represent the observables.

For each atom $\bar{p}$ we get a state $\mu_{\bar{p}}$ on $L$ by
$\mu_{\bar{p}}(\bar{q}) := \left|\left\langle p|q\right\rangle\right| \leq 1$ and \linebreak
$\mu_{\bar{p}}(\bar{q}^{\bot}) := 1 - \left|\left\langle p|q\right\rangle\right|$
for the points $\bar{q}$ and the lines $\bar{q}^{\bot}$. 
Of course $\mu_{\bar{p}}(0):=0$ and $\mu_{\bar{p}}(\mathbb{I}):=1$.
The convex hull of $\left\{\mu_{\bar{p}} | \bar{p} \in (\mathbb{C} \otimes \mathbb{O})\mathbb{P}^{2}\right\}$ 
becomes the state space~$S$. Then
$ \mathbb{P}(v|\bar{p}) = \mu_{\bar{p}}(v) $ for any atom $\bar{p}$ and any $v \in L$
and the transition probability between two atoms $\bar{p}$ and $\bar{q}$ is
$ \mathbb{P}(\bar{q}|\bar{p}) = \left|\left\langle q|p\right\rangle\right| $.
A convex metric on the compact space $(\mathbb{C} \otimes \mathbb{O})\mathbb{P}^{2}$ is
$$d(p,q) := arccos( \sqrt{\left|\left\langle p|q\right\rangle\right|} ) 
= arccos( \sqrt{\mathbb{P}(q|p)} ).$$
This metric is not normalized, but $(2/\pi) d$ is.
The transition probability is not homogeneous, since that would entail
that the above metric becomes two-point homogeneous, which is ruled out by Theorem~2.
An interesting feature, however, is the $E_6$-symmetry of
this transition probability. Moreover, $L$ contains many 
subsystems that possess a more satisfying structure 
and are isomorphic to the the quantum logics 
in $H_3(\mathbb{R})$, $H_3(\mathbb{C})$, $H_3(\mathbb{H})$ 
and $H_3(\mathbb{O})$.

Further projective planes in the wider sense are 
$(\mathbb{H} \otimes \mathbb{O})\mathbb{P}^{2}$ and 
$(\mathbb{O} \otimes \mathbb{O})\mathbb{P}^{2}$
with the exceptional symmetry groups $E_7$ and $E_8$ \cite{baez2002octonions}. 
The Cartan labels of the corresponding symmetric spaces 
are $EVI$ and $EVIII$. It is not known
whether the above construction of a quantum logic, a state space
and the transition probabilities is possible in these cases as well.

\section{Conclusions}

We have seen that the quantum theoretical transition probability
features a maximum degree of homogeneity.
Not only the complex matrix algebras
which represent the common model of quantum mechanics
possess this feature, but all other simple Euclidean Jordan algebras 
do as well.

Well-established mathematical accomplishments 
due to Wang and Hirzebruch yield that 
the atomic parts or, equivalently, the pure state spaces of the simple
Euclidean Jordan algebras can be characterized in a very basic way 
by purely topological means.
This is an important difference to many other more recent approaches 
that aim to distinguish the entire state
spaces among the convex compact sets~\cite{AS02, 
BarnumHilgert2020, mielnik1968geometry, 
muller2012ududec, nie2023conv_self-dual}.
In searching for the origin of quantum probability we 
arrive at a purely topological structure. 

The homogeneity of the common quantum theoretical transition probability 
is not restricted to its finite-dimensional version,
but Theorem~1 and Theorem~2 require a finite dimension. 
Unfortunately, an extension of the topological characterization 
of the system of atoms or of the extreme boundary of the state space
(the system of the pure states) in such a way that
infinite-dimensional quantum theory gets included, is not at hand,
since compactness requires weak topologies 
that cannot be defined by metrics then. Moreover, in
infinite-dimensional quantum theory, the von Neumann factors 
of the types II and III play an important role, but 
do not contain any atom.

Besides the reducible (non-simple) cases we have identified a further 
irreducible case with non-homogeneous transition probability. This is a 
rather exotic model. The bioctonionic projective plane 
becomes the quantum logic and the atoms form the symmetric space $EIII$;
its isometry group is the exceptional Lie group $E_6$. 
This model is by far less
pleasant than the well-understood quantum logics and state spaces of the Euclidean 
Jordan algebras with their rich mathematical structure, but features the exceptional
$E_6$-symmetry, which is sometimes considered a candidate 
for internal symmetries in particle physics~\cite{catto2003, gursey1978octonionic}.
Though the overall structure of this model appears rather weak,
it contains subsystems with the more satisfying 
and better understood structures of the quantum logics 
of the matrix algebras $H_3(\mathbb{R})$, $H_3(\mathbb{C})$, 
$H_3(\mathbb{H})$ and~$H_3(\mathbb{O})$ in many ways.\\

\footnotesize
\noindent 
\textbf{Funding information} - not applicable\\
\textbf{Data availability} - not applicable\\
\textbf{Conflict of interest} - not applicable
\small
\bibliographystyle{abbrv}
\bibliography{Literatur2025}
\end{document}